\documentclass[a4paper]{jpconf}

\usepackage[english]{babel}
\usepackage{graphicx}
\usepackage{color}
\usepackage{amssymb}
\usepackage{amsmath}
\usepackage{enumitem}
\usepackage{float}

\newcommand{\Elab}{E_\mathrm{lab}}

\begin{document}

\title{Cooper-Frye Negative Contributions in a Coarse-Grained
  Transport Approach}

\author{D.~Oliinychenko$^{1,3}$, P.~Huovinen$^{1,2}$, H.~Petersen$^{1,2}$}

\address{$^1$Frankfurt Institute for Advanced Studies, D-60438
  Frankfurt am Main, Germany}
\address{$^2$Institut f\"ur Theoretische Physik, Goethe-Universit\"at,
  D-60438 Frankfurt am Main, Germany}
\address{$^3$Bogolyubov Institute for Theoretical Physics, Kiev 03680, Ukraine}

\ead{oliiny@fias.uni-frankfurt.de}

\begin{abstract}
  Many models of heavy ion collisions employ relativistic
  hydrodynamics to describe the system evolution at high
  densities. The Cooper-Frye formula is applied in most of these
  models to turn the hydrodynamical fields into particles. However, the number
  of particles obtained from the Cooper-Frye formula is not always
  positive-definite. Physically negative contributions of the
  Cooper-Frye formula are particles that stream backwards into the
  hydrodynamical region.

  We quantify the Cooper-Frye negative contributions in a
  coarse-grained transport approach, which allows to compare them to
  the actual number of underlying particles crossing the transition
  hypersurface. It is found that the number of underlying inward
  crossings is much smaller than the one the Cooper-Frye formula gives under the
  assumption of equilibrium distribution functions. The magnitude of Cooper-Frye
  negative contributions is also investigated as a function of hadron
  mass, collision energy in the range $E_{\rm lab} = 5-160A$ GeV, and
  collision centrality. The largest negative contributions we find are
  around 13\% for the pion yield at midrapidity at $E_{\rm lab} = 20A$ GeV
  collisions.
\end{abstract}

\section{Introduction}

In the hot and dense system of strongly interacting matter created in
heavy ion collisions the mean free path of the particles is much
smaller than the size of the fireball. This fact together with the
assumption of fast thermal equilibration allows to apply relativistic
hydrodynamics for the dynamical description of heavy ion
collisions. At later times of the evolution and at its edges the system is
dilute, the mean free path is larger than system size, and
hydrodynamics is not applicable. An adequate description of the system
in those circumstances is provided by kinetic (transport) equations,
such as the Boltzmann equation or its modifications for the quantum
case. State of the art simulations of heavy ion collisions couple
hydrodynamics for the early stage of the evolution to hadron transport
for the late stage. Such approaches are called hybrid
approaches~\cite{Hirano:2012kj}.

The transition from hydrodynamics to transport (so-called particlization)
is non-trivial, because hydrodynamics contains no microscopic
information that is needed for transport. Currently, most of the models
implement particlization in the following way. Hydrodynamical
equations are solved in the whole forward light cone, including those
regions, where hydrodynamics is not applicable. Then the particlization
hypersurface is found from the hydrodynamical evolution. Usually a
hypersurface of constant time, temperature or energy density is
taken. Particle distributions to be fed into the kinetic model are
generated on this hypersurface according to the Cooper-Frye
formula~\cite{ref:CF74}:
\begin{eqnarray}
p^0 \frac{dN}{d^3p} = p^{\mu} d\sigma_{\mu} f(p)
\end{eqnarray}
Here $\frac{dN}{d^3p}$ is a spectrum of particles emerging from an
element of the hypersurface, $d\sigma_{\mu}$ is the normal four-vector to
the hypersurface, its length being equal to the area of hypersurface
element. In ideal fluid calculations the distribution function $f(p)$ is
taken as an equilibrium Fermi or Bose distribution: $f(p) = \left[
  exp\left(\frac{p^{\mu}u_{\mu} - \mu}{T} \right) \pm 1 \right]^{-1}$.
An advantage of the Cooper-Frye formula is that it respects
conservation laws. The disadvantage is that for space-like elements of
the hypersurface (\emph{i.e.} where $d\sigma^{\mu}d\sigma_{\mu} < 0$)
there exist particle momenta $p^{\mu}$ such that 
$p^{\mu}d\sigma_{\mu} < 0 $ and thus $\frac{dN}{d^3p} < 0$ in the
Cooper-Frye formula. This is clearly unphysical, because the number of
particles must be positive in every element of phase space. However,
in all practical cases after integration over the hypersurface one
gets positive definite spectra. One can split the integrated
Cooper-Frye formula into positive and negative parts:
\begin{eqnarray}
\frac{dN}{d^3p} = \int_{\sigma} \frac{p^{\mu}}{p^0} d\sigma_{\mu} f(p) \Theta(p^{\mu}d\sigma_{\mu}) +
                    \int_{\sigma} \frac{p^{\mu}}{p^0} d\sigma_{\mu} f(p) \Theta(-p^{\mu}d\sigma_{\mu}) \,.
\end{eqnarray}
The second term is called Cooper-Frye negative contributions and the
first one - positive. For the applicability of the Cooper-Frye formula
the negative contributions must be much smaller than the positive ones. This is
usually true for simulations of high energy collisions. Negative
contributions of around 9\% relative to the positive ones were reported for RHIC
top energy~\cite{ref:PetHuo12}. At the same time negative
contributions at $\Elab = 160 A$GeV were found to be around 13\%. This
suggests that at lower collision energies negative contributions might
become large and Cooper-Frye formula will be inapplicable. Does it
really happen? To answer this question we systematically investigate
the negative Cooper-Frye contributions differentially in rapidity and
transverse momentum against collision energy, centrality and particle
species.

Several ways to circumvent the problem of negative contributions were
suggested (see~ \cite{Oliinychenko:2014tqa} for a short summary). One
practical way to do it in hybrid models would be to simultaneously
neglect negative contributions and remove the particles from the transport
calculation, if they fly to the hydrodynamic region. If the distributions in
hydrodynamics and cascade are the same, then conservation laws will be
fulfilled. Such an approach has never been implemented. To check if it is
feasible, we compare negative contributions from the Cooper-Frye formula to
distributions of backscattered particles from the cascade. We take
advantage of the coarse-grained approach, which allows to calculate
both within one framework.

\section{Methodology}

We aim at two goals: systematic estimation of Cooper-Frye negative
contributions and comparing them to distributions of transport
particles flying inwards to the hydrodynamics region. As a transport
model we take Ultra-Relativistic Quantum Molecular Dynamics (UrQMD
3.3p2)~\cite{ref:UrQMD}. UrQMD allows to simulate heavy ion collisions
as a sequence of elementary particle collisions. Included processes
are 2 to 2 scattering, resonance formation and decays, string
excitation and fragmentation.

We generate an ensemble of UrQMD Au+Au collision events and average
them on a rectangular grid to obtain the energy momentum tensor
$T^{\mu\nu}$ and the baryon current $j^{\mu}$ in each cell. In each cell
we find the Landau rest frame (LRF, a frame, where energy flow is zero:
$T_{LRF}^{0i} = 0$). We obtain the energy density
$\epsilon_{LRF} = T^{00}_{LRF}$, flow velocity $u^{\mu}$ and baryon
density in the LRF $n_{LRF} = j^{\mu}_B u_{\mu}$. Knowing $\epsilon_{LRF}$
for each grid cell we construct the hypersurface $\Sigma$ of constant
$\epsilon_{LRF}$ and find the normal vectors $d\sigma_{\mu}$ for each piece of
$\Sigma$. The latter is done using the Cornelius
subroutine~\cite{ref:PetHuo12}, that provides a continuous surface
without holes and avoids double counting of hypersurface pieces. The
hypersurface $\Sigma$ mimics the transition hypersurface in hybrid
models. When $\Sigma$ is obtained we perform a Cooper-Frye calculation
on it and compare to distributions of underlying UrQMD particles that
cross $\Sigma$.

We perform our calculations with time step $\Delta t$ = 0.1 fm/c, grid
spacing in the beam direction $\Delta z = $0.3 fm, and grid spacings
in transverse direction $\Delta x = \Delta y$ = 1 fm. For collision
energy $\Elab = 160 A$ GeV we take $\Delta z = $0.1 fm, and
$\Delta x = \Delta y$ = 0.3 fm. We have checked that for such a choice
of grid spacing conservation laws on the surface are fulfilled with
an accuracy better than 1\%. In other words 
$\int_{\Sigma} T^{\mu 0} d\sigma_{\mu}$ and total energy flowing out
of the hypersurface calculated by particles differ by no more than 1\%. To
create a smooth hypersurface and obtain reproducible results we employ 
a Gaussian smearing procedure. For the construction of the hypersurface every UrQMD
particle with coordinates $(t_p,x_p,y_p,z_p)$ and 4-momentum $p^{\mu}$
is substituted by 300 marker particles with coordinates distributed
with the probability density
$f(x,y,z) \sim \exp \left(-\frac{(x-x_p)^2}{2\sigma^2} 
                  - \frac{(y-y_p)^2}{2\sigma^2}
                  - \gamma_z^2 \frac{(z-z_p)^2}{2\sigma^2}\right)$, 
where $\gamma_z = (1-p^z/p^0)^{-1/2}$. In this way every particle
contributes to $T^{\mu\nu}$ and $j^{\mu}$ not only of the cell, where
it is located, but also to the adjacent cells. We take as number of
events $N=1500$ and the Gaussian width $\sigma = 1$ fm. Choice of $N$,
$\sigma$, grid spacing and sensitivity of results to these choices are
discussed in~\cite{Oliinychenko:2014tqa}.

\section{Results}

If the distribution of UrQMD particles is thermal on some closed
hypersurface and the system is in chemical equilibrium, then the 
Cooper-Frye formula should give results identical to explicit particle
counting. This would allow to compensate negative contributions by
removing particles from UrQMD, if they cross the hypersurface inwards. In
such a treatment conservation laws on the surface would be respected. We
check, if this can be done on the hypersurface $\Sigma$ of constant
energy density in Landau frame, $\epsilon = 0.3$ GeV/fm$^3$.

\noindent
 \begin{minipage}{\linewidth}
  
      \centering
      \begin{minipage}{0.61\linewidth}
       \begin{figure}[H]
       \centering
       \vspace{-0.6cm}
       \includegraphics[height=4.8cm]{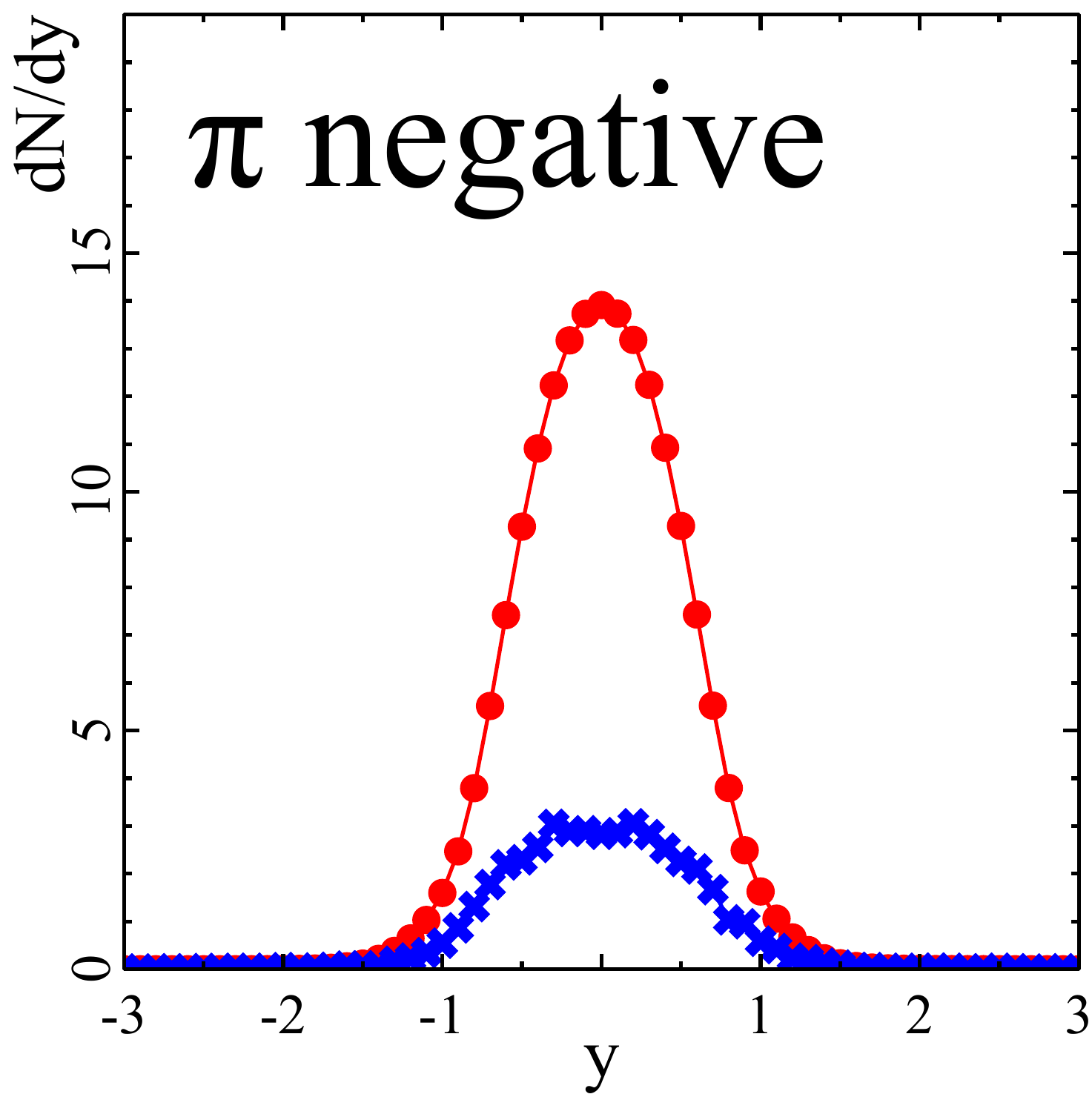}
       \includegraphics[height=4.8cm]{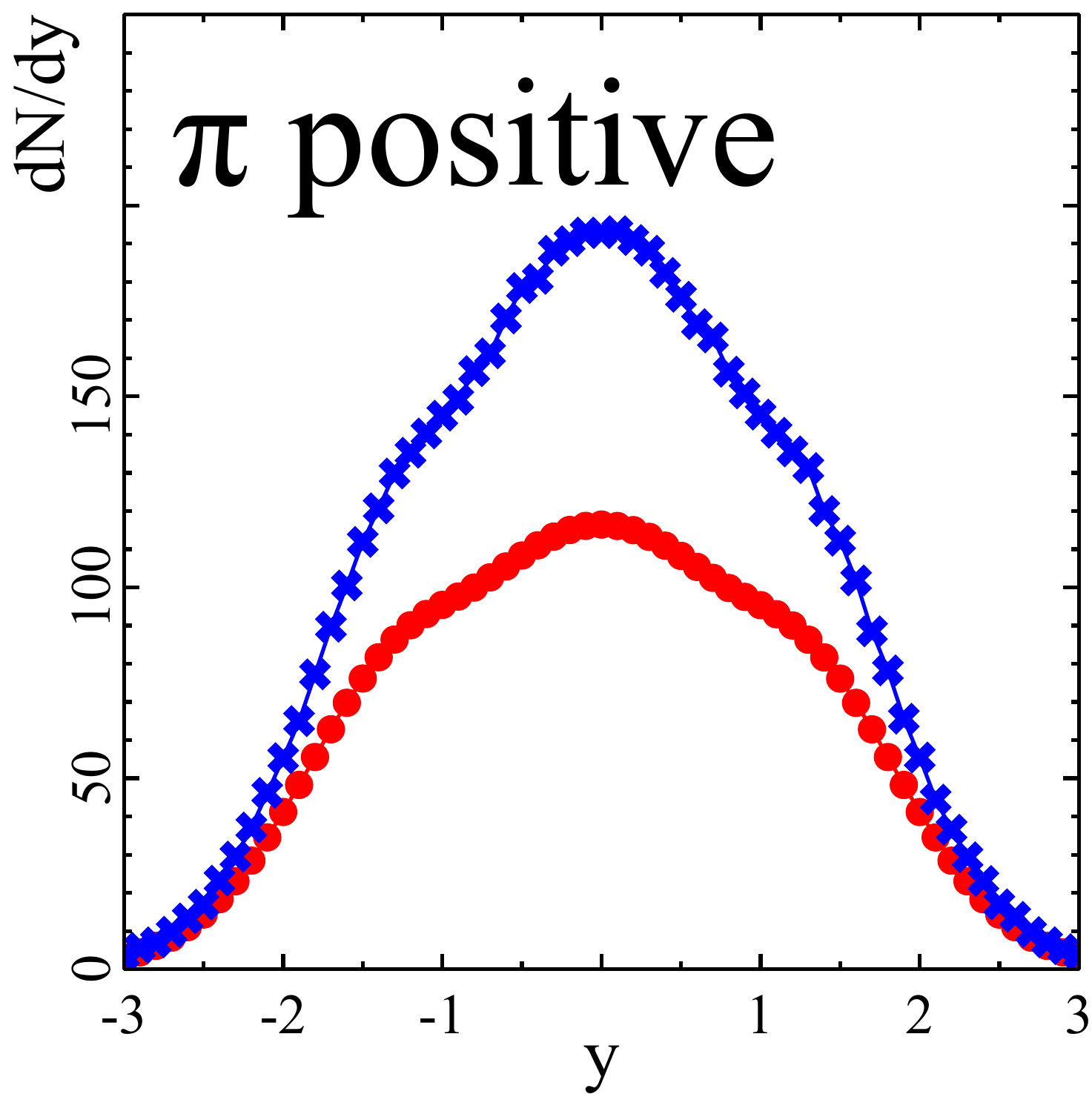}
       \caption{Cooper-Frye rapidity spectra for pions on $\Sigma$
         (red circles) are compared to distribution of UrQMD pions
         crossing $\Sigma$ (blue crosses). Left panel: negative
         contributions and inward crossings. Right panel: positive
         contributions and outward crossings. Collision energy $\Elab
         = 40$~$A$GeV, central collisions. Note the very different
         scale of negative and positive contributions.}
       \label{Fig:pos_contr_rap}
       \end{figure}
      \end{minipage}
      \hfill
      \begin{minipage}{0.32\linewidth}
        \begin{figure}[H]
        \centering
       \includegraphics[height=4.8cm]{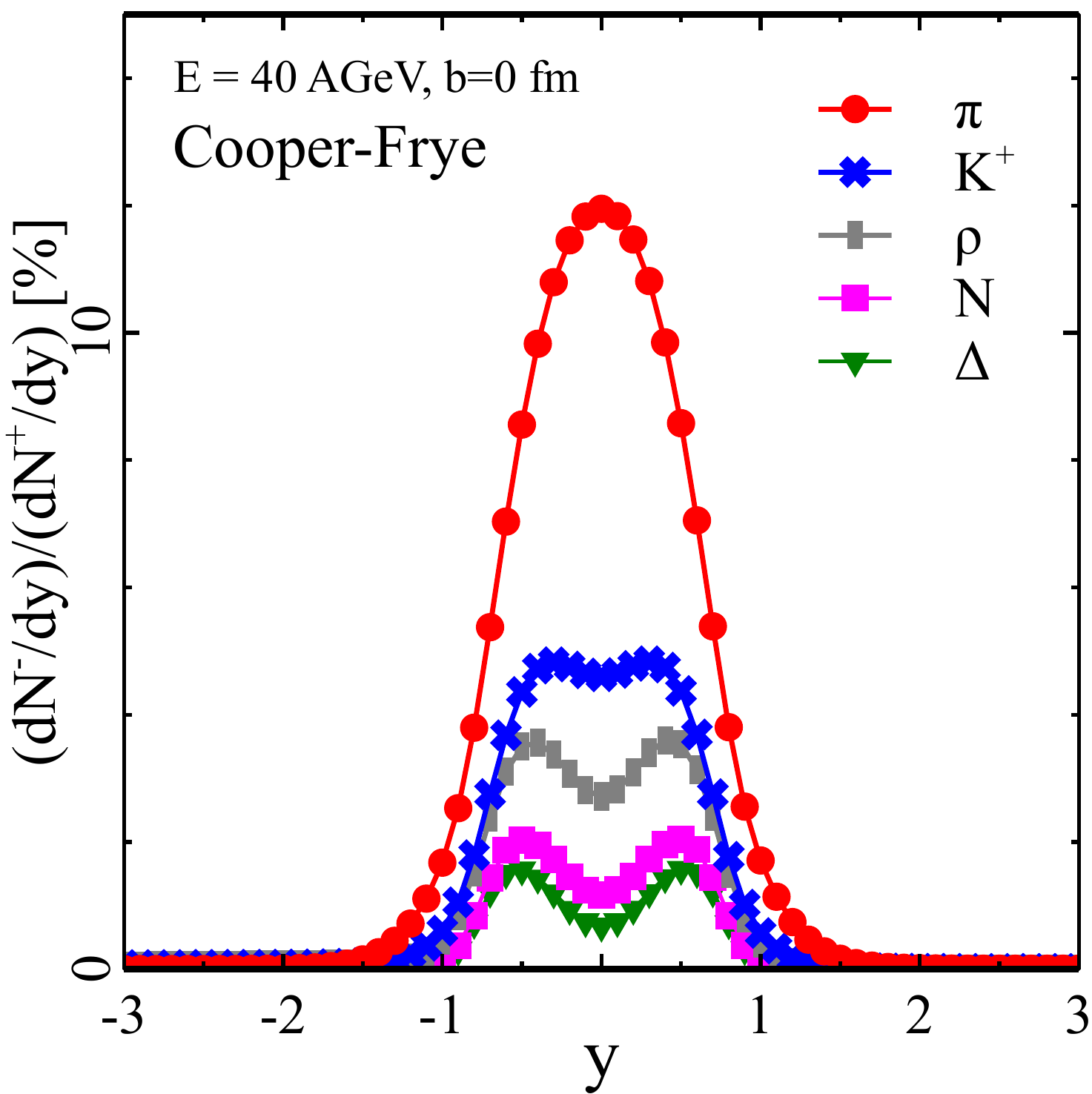}
       \caption{\small Rapidity distribution of the ratio of negative
         to positive contributions for different hadrons: pions
         (red circles), $K^{+}$ (blue crosses), $\rho$ (grey bars),
         nucleons (magenta rectangles) and $\Delta$ (green triangles).
         Cooper-Frye calculation in central Au+Au collisions at $\Elab
         = 40\ A$GeV.}
       \label{Fig:neg_contr_pmass}
        \end{figure}
      \end{minipage}
      \vspace{0.3cm}
 \end{minipage}

When calculating the net number of pions passing through the surface,
one finds that the number of pions in UrQMD is larger than in the
equilibrated Cooper-Frye calculation~\cite{Oliinychenko:2014tqa}. It
might be possible to explain this as a sign of chemical
non-equilibrium in UrQMD, but when one looks at the positive and
negative contributions to the pion distributions shown in
Fig.~\ref{Fig:pos_contr_rap}, one sees that a difference in the pion
density only is not sufficient to explain the differences in the 
contributions. The positive contribution is much larger in UrQMD than
in the Cooper-Frye calculation, whereas the negative contribution depicts the
opposite behaviour: in UrQMD it is much smaller than in
the Cooper-Frye scenario. This kind of distributions may indicate that the
collective flow velocity of pions is much larger than the collective
velocity of other particles~\cite{Sorge:1995pw,Pratt:1998gt}, or that
the dissipative corrections are very large.

Since neither negative contributions coincide with inward crossings,
nor positive contributions coincide with outward crossings, the above
mentioned idea of compensating negative contributions will not work
in our case. Instead we concentrate on finding when negative
contributions play the most prominent role, and evaluate the ratio of
negative contributions to positive ones to estimate the error they bring
into hybrid calculations. For that we vary hadron sort, collision energy and
centrality. As a relevant variable we consider the ratio of negative
to positive contributions integrated over the hypersurface,
$(dN^-/dy)/(dN^+/dy)$.
  
From Fig.~\ref{Fig:neg_contr_pmass} one can see that negative
contributions become smaller for larger particle mass. It is simple to
understand this result in the rest frame of the fluid element where
the surface is moving. If the hypersurface $\Sigma$ moves inwards, as
is usually the case in fluid-dynamical calculations, a particle must
be faster than the surface to cross it inwards. For larger particle
mass the average velocities of the thermal motion are smaller and the probability
to catch up the hypersurface is also smaller. Therefore, to find
maximal negative contributions it is sufficient to consider pions
only. For them we find out that if negative contributions are binned
according to rapidity, then the largest contribution occurs at
midrapidity.

We show the negative to positive contribution ratio for the pion yield at
midrapidity for collision energies $\Elab = 5$--$160 A$ GeV in
Fig.~\ref{Fig:neg_contr_Ecoll}. The ratio in UrQMD calculation is much
smaller than in the equilibrated Cooper-Frye calculation at all collision
energies, but the maximum lies in the region of $20 A$ GeV in
both approaches. The value of the maximum is about 13\% in Cooper-Frye
calculation, and about 4\% in UrQMD calculation.
     
\noindent
 \begin{minipage}{\linewidth}
  
      \centering
      \begin{minipage}{0.48\linewidth}
       \begin{figure}[H]
       \centering
       \vspace{-0.2cm}
      \includegraphics[height=5.2cm]{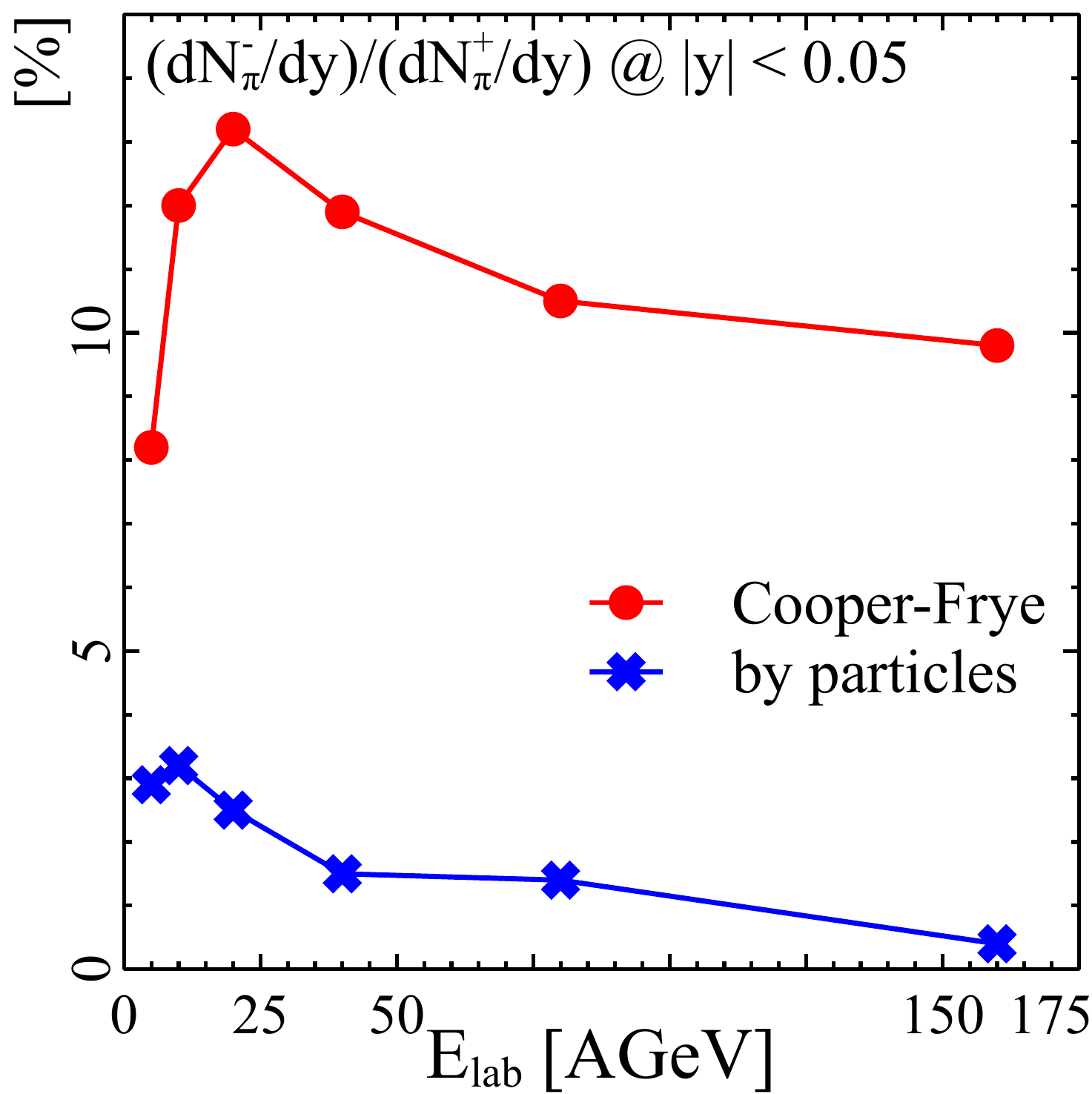}
       \caption{\small The ratio of negative to positive contributions
         on the $\epsilon(t,x,y,z) = \epsilon_c = 0.3$ GeV/fm$^3$
         surface for pions at midrapidity. Red circles depict the
         ratio in a Cooper-Frye calculation assuming thermal
         equilibrium, and blue crosses in explicit calculation of
         UrQMD particles.}
       \label{Fig:neg_contr_Ecoll}       
       \end{figure}
      \end{minipage}
      \hspace{0.02\linewidth}
      \begin{minipage}{0.48\linewidth}
        \begin{figure}[H]
        \centering
        \vspace{-0.6cm}
        \includegraphics[height=5.2cm]{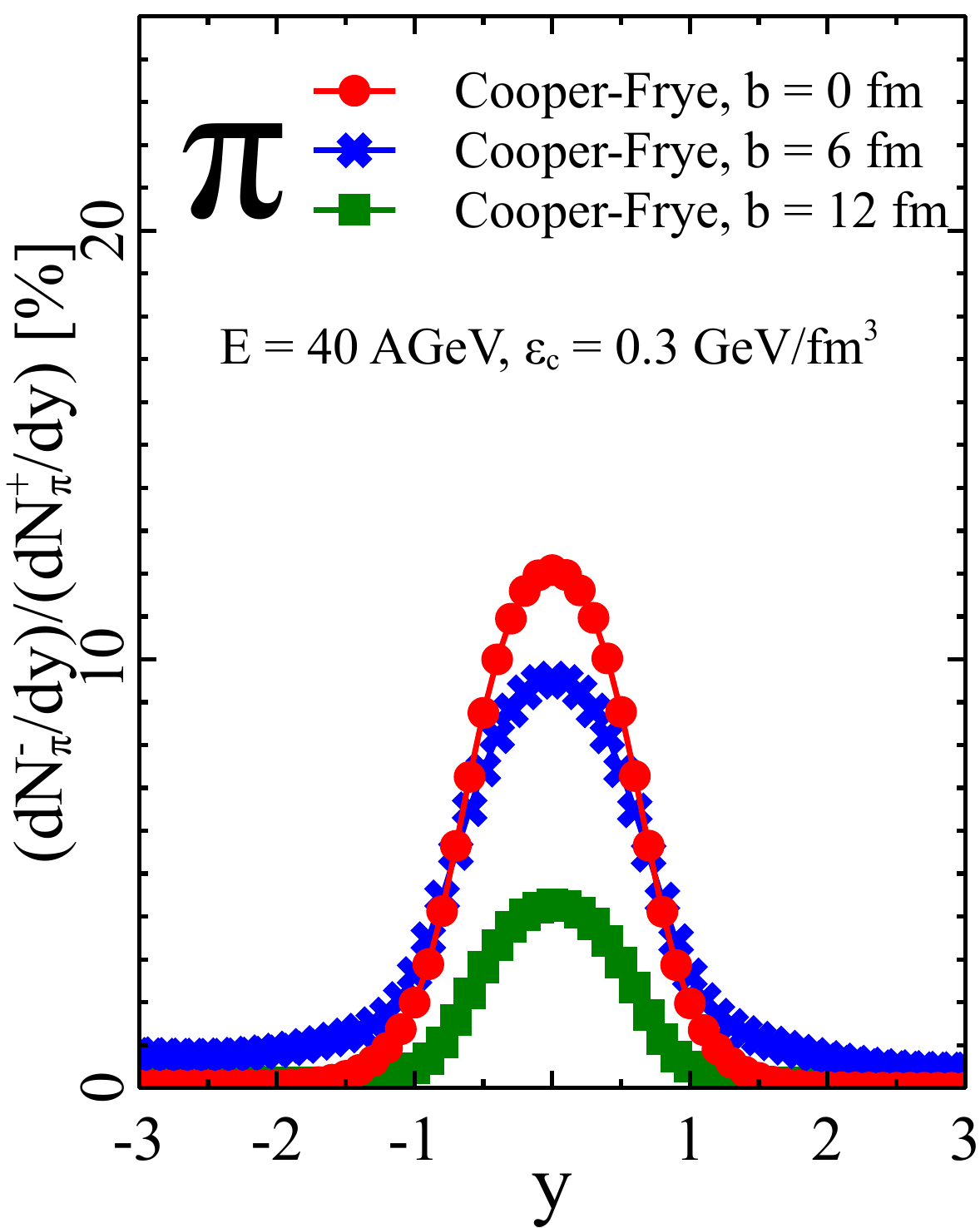}
        \caption{\small Rapidity distribution of the ratio of negative to
          positive contributions for pions in Au+Au collisions at 
          $\Elab = 40\ A$GeV at various centralities: $b=0$ (red circles), 
          $b = 6$ fm (blue crosses) and $b = 12$ fm (green rectangles).}
        \label{Fig:neg_contr_b}
        \end{figure}
      \end{minipage}
      \vspace{0.3cm}
 \end{minipage}

The dependency of negative to positive contribution ratio on collision
energy is non-monotonous. This is not obvious, because several factors
influence the ratio: the temperature on the hypersurface, the relative velocities
between the flow and the surface, and the relative amounts of volume and surface
emission, \emph{i.e.} emission from the time- and space-like parts of
the surface. Larger temperature results in larger negative
contributions because the thermal velocities increase. Larger relative
velocity leads to smaller negative contributions. The larger the
relative amount of volume-emission, the smaller the negative
contributions. With increasing collision energies the temperature
saturates and thus the changes in the last two factors make negative
contributions fall with increasing collision energy. The decrease of
negative contributions at higher energy is predictable, because
the relative amount of volume emission and relative velocity between flow
and surface increase with collision energy. However, the behaviour of
negative contributions at lower energy is caused by the interplay of
all three factors.

We plot negative contributions for pions versus collision centrality
in Fig.~\ref{Fig:neg_contr_b}, and find that for peripheral collisions
negative contributions are smaller than for central collisions. The relative
amount of surface and volume emission plays the most prominent role
here. For peripheral collisions volume emission dominates due to
the short lifetime of the system and negative contributions are
small.

After varying the collision energy and the centrality, we can conclude that in
the worst scenario negative contributions can hardly exceed
15\%. Changing the criterion of hypersurface to $\epsilon_c = 0.6$~GeV/fm$^3$
does not change this conclusion (see Ref.~\cite{Oliinychenko:2014tqa}).

\section{Conclusions}

We have investigated negative Cooper-Frye contributions and
backscattering using a coarse-grained molecular dynamics
approach. Au+Au collisions at $\Elab = 5$--$160\ A$ GeV energies have
been simulated using UrQMD, and a hypersurface $\Sigma$ of constant
Landau rest frame energy density has been constructed. On this surface
we have calculated two quantities: The ratio of Cooper-Frye negative
to positive contributions ($r_{eq}$), which assumes local thermal
equilibrium, and the ratio of UrQMD particles crossing $\Sigma$ inward
to crossing $\Sigma$ outward ($r_{neq}$), which does not assume
equilibrium.

We found that at all collision energies $r_{eq} \gg r_{neq}$. We
explain this by a deviation of pions in UrQMD simulation from
equilibrium. A non-monotonous dependency of $r_{eq}$ and $r_{neq}$ on
collision energy was found with a maximum at 10-20 $A$ GeV, maximal
$r_{eq}$ being around 13\%. The size of the negative contributions is a
result of an interplay of several factors: the temperature on the
hypersurface, the relative velocities between flow and surface, and
the relative amounts of volume and surface emission.

\ack
  This work was supported by the Helmholtz International Center for
  the Facility for Antiproton and Ion Research (HIC for FAIR) within
  the framework of the Landes-Offensive zur Entwicklung
  Wissenschaftlich-Oekonomischer Exzellenz (LOEWE) program launched by
  the State of Hesse. DO and HP acknowledge funding of a Helmholtz
  Young Investigator Group VH-NG-822 from the Helmholtz Association
  and GSI, and PH by BMBF under contract no.~06FY9092. DO acknowledges
  support of HGS-HIRe. Computational resources have been provided by
  the Center for Scientific Computing (CSC) at the Goethe University
  of Frankfurt.


\section*{References}

\begin{thebibliography}{99}

\bibitem{Hirano:2012kj} 
  T.~Hirano, P.~Huovinen, K.~Murase and Y.~Nara,
  Prog.\ Part.\ Nucl.\ Phys.\  {\bf 70}, 108 (2013).


\bibitem{ref:CF74}
F.~Cooper and G.~Frye, Phys.\ Rev.\ D {\bf 10}, 186 (1974).
nd W. Greiner, Phys. Rev. C 59 (1999) 1

\bibitem{ref:UrQMD}
S.~A.~Bass, M.~Belkacem et al., Prog.\ Part.\ Nucl.\ Phys.\ {\bf 41}, 225
(1998);
\nonum M.~Bleicher, E.~Zabrodin, et al., J.~Phys.~G: Nucl.\ Part.\
Phys.\ {\bf 25}, 1859 (1999).

\bibitem{ref:PetHuo12}
P. Huovinen, H. Petersen, 
Eur.~Phys.~J.~A {\bf 48}, 171 (2012).

\bibitem{Oliinychenko:2014tqa}
  D.~Oliinychenko, P.~Huovinen and H.~Petersen,
  arXiv:1411.3912 [nucl-th].

\bibitem{Sorge:1995pw}
  H.~Sorge,
  Phys.\ Lett.\ B {\bf 373}, 16 (1996).

\bibitem{Pratt:1998gt}
  S.~Pratt and J.~Murray,
  Phys.\ Rev.\ C {\bf 57}, 1907 (1998).



\end{thebibliography}
\end{document}